\RequirePackage{silence}
\documentclass[aps,pra,twocolumn,showpacs,amsmath,amssymb,preprintnumbers,superscriptaddress,10pt,longbibliography]{revtex4-2}

\usepackage{graphicx}
\graphicspath{{figures/}}
\usepackage{SIunits}
\usepackage{hyphenat}
\usepackage[bookmarks=false]{hyperref}
\usepackage{color}
\usepackage[capitalise]{cleveref}
\crefname{section}{Sec.}{Secs.}
\Crefname{section}{Section}{Sections}
\usepackage{multirow}
\usepackage{makecell}

\definecolor{red}{RGB}{17,84,143}
\definecolor{pink}{RGB}{255,0,255}

\definecolor{red}{rgb}{1,0,0}

\definecolor{blue}{RGB}{0,212,17}

\definecolor{amethyst}{RGB}{126, 32, 203}

\definecolor{darkgreen}{RGB}{0, 100, 0}

\definecolor{orange}{RGB}{255,179,92}

\begin{document}
\title{Pulsed laser attack at 1061~nm potentially compromises quantum key distribution}

\author{Anastasiya~Ponosova}
\email{nastya-aleksi@mail.ru}
\affiliation{Russian Quantum Center, Skolkovo, Moscow 121205, Russia}
\affiliation{NTI Center for Quantum Communications, National University of Science and Technology MISiS, Moscow 119049, Russia}
\affiliation{Prokhorov General Physics Institute of Russian Academy of Sciences, Moscow 119991, Russia}

\author{Irina~Zhluktova}
\affiliation{Prokhorov General Physics Institute of Russian Academy of Sciences, Moscow 119991, Russia}

\author{Daria~Ruzhitskaya}
\affiliation{Russian Quantum Center, Skolkovo, Moscow 121205, Russia}
\affiliation{NTI Center for Quantum Communications, National University of Science and Technology MISiS, Moscow 119049, Russia}

\author{Daniil~Trefilov}
\affiliation{Vigo Quantum Communication Center, University of Vigo, Vigo E-36310, Spain}
\affiliation{School of Telecommunication Engineering, Department of Signal Theory and Communications, University of Vigo, Vigo E-36310, Spain}
\affiliation{atlanTTic Research Center, University of Vigo, Vigo E-36310, Spain}
\affiliation{Russian Quantum Center, Skolkovo, Moscow 121205, Russia}
\affiliation{National Research University Higher School of Economics, Moscow 101000, Russia}

\author{Anqi~Huang}
\affiliation{College of Computer Science and Technology, National University of Defense Technology, Changsha 410073, China}

\author{Alexey~Wolf}
\affiliation{Institute of Automation and Electrometry of Siberian Branch of RAS, Novosibirsk, 630090, Russia}

\author{Vladimir~Kamynin}
\affiliation{Prokhorov General Physics Institute of Russian Academy of Sciences, Moscow 119991, Russia}

\author{Vladimir~Tsvetkov}
\affiliation{Prokhorov General Physics Institute of Russian Academy of Sciences, Moscow 119991, Russia}

\author{Vadim~Makarov}
\affiliation{Russian Quantum Center, Skolkovo, Moscow 121205, Russia}
\affiliation{Vigo Quantum Communication Center, University of Vigo, Vigo E-36310, Spain}
\affiliation{NTI Center for Quantum Communications, National University of Science and Technology MISiS, Moscow 119049, Russia}

\date{\today}

\begin{abstract}
Quantum key distribution systems offer cryptographic security, provided that all their components are thoroughly characterised. However, certain components might be vulnerable to a laser-damage attack, particularly when attacked at previously untested laser parameters. Here we show that exposing $1550$-$\nano\meter$ fiber-optic isolators to $1061$-$\nano\meter$ sub-nanosecond pulsed illumination with $1.16~\watt$ average power permanently degrades their isolation at $1550~\nano\meter$, while their forward transparency is less affected. One experimental sample was exposed to $17$-$\milli\watt$ average power picosecond attacking pulses that temporarily reduced its isolation below the specified guaranteed minimum value. This indicates a potential security threat in these attacking laser regimes that need to be addressed by improving security analysis for various light-injection attacks.
\end{abstract}

\maketitle


Quantum key distribution (QKD) systems are vital for secure communications in the coming era of quantum computers, as they use quantum physics to generate cryptographic keys resistant to computational attacks~\cite{lo2014, gisin2002}. Despite robust theoretical security proofs~\cite{lo1999,shor2000,koashi2009}, practical implementations remain vulnerable to side-channel attacks arising from device imperfections~\cite{lo2014,sun2022,marquardt2023,makarov2024}. To date, researchers have demonstrated nearly 30 hacking strategies targeting QKD apparatus. These strategies can be divided based on the vulnerable system part: transmitter~\cite{brassard2000,fung2007,lucamarini2015, huang2019,lovic2023,ye2023,lu2023,tan2025,fadeev2025}, receiver~\cite{vakhitov2001, makarov2005, gisin2006, makarov2006,weier2011,lydersen2011b,meda2017,acheva2023}, and flaws in the calibration process~\cite{jain2011,fei2018}.
One such attack on a transmitter is the laser-damage attack (LDA), which targets optical components to change their operational characteristics by injection of high-power laser emission~\cite{bugge2014, makarov2016,huang2020, ruzhitskaya2021,ponosova2022}. 
Existing studies have focused mainly on LDAs using continuous-wave (cw) lasers at $1550~\nano\meter$, the common telecom wavelength~\cite{lucamarini2015,makarov2016,huang2020, ponosova2022,bugai2022,peng2024}. Their findings suggest the integration of protective countermeasures such as isolators~\cite{huang2020, ponosova2022}, power limiters~\cite{peng2024} and monitoring systems~\cite{lucamarini2015, dixon2017}. However, these studies do not consider other possibilities for LDA, such as the use of pulsed lasers (PL) and lasers with other than $1550$-$\nano\meter$ wavelength. Recent studies show that fiber-optic components designed to protect QKD systems against LDA by 1550-nm cw lasers may be vulnerable to attacks by PL at various operating wavelengths~\cite{ruzhitskaya2021, ruzhitskaya2023, kang2025, tsypkin2024}. Addressing this gap is crucial, as the current QKD security analysis~\cite{sajeed2021,makarov2024} lacks empirical validation against such scenarios of LDA, potentially leaving exploitable loopholes for an eavesdropper.


\begin{figure*}
\centering
	\includegraphics{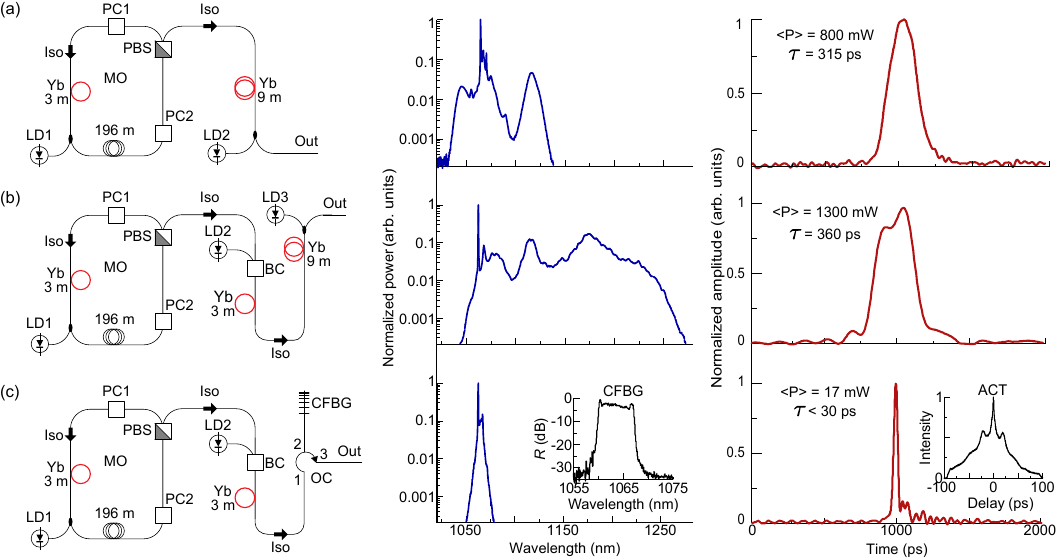}
  \caption{Pulsed laser schemes, output spectra, and pulse shapes. (a)~Single-YDFA scheme. (b)~Dual-YDFA scheme.
  (c)~Pulse-compression scheme. BC, beam combiner; LD, laser diode; ISO, isolator; PC, polarization controller; PBS, polarization beamsplitter; CFBG, chirped fiber Bragg grating; MO, master oscillator. Black ellipses depict splices of components with distributed side-coupled cladding-pumped (DSCCP) Yb-doped fibers (so-called GTWave fiber) that have one active, Yb-doped fiber and one passive fiber within a common silicone coating. Insets show the reflection spectrum of CFBG and the autocorrelation trace (ACT). Spectra and oscillograms are normalized to their peak value.}
  \label{fig:PL}
\end{figure*}

Based on our previous research demonstrating the possibility of LDA using PL~\cite{ruzhitskaya2021, ruzhitskaya2023}, here we expose fiber-optic isolators from QKD systems to 1061-$\nano\meter$ pulsed laser. In principle, an eavesdropper is free to select any laser tool for the hacking. In our research, we prefer $1061$-$\nano\meter$ lasers because high-power ytterbium-doped single-mode fiber lasers are widely available as commercial products~\cite{ipg2025}. They are readily accessible tools that significantly increase the feasibility of attacks. An additional advantage of this wavelength for hacking is that protective fiber-optic isolators are known to have a dip in isolation, most pronounced over the wavelength range from about 1050 to $1300~\nano\meter$~\cite{jain2015,ruzhitskaya2021,tan2025}.
We employ a configurable fiber laser system to generate sub-nanosecond (200--$400~\pico\second$) and picosecond (less than $30~\pico\second$) pulses at $1061~\nano\meter$, with average output power of $1300~\milli\watt$ and $17~\milli\watt$, respectively. We test isolators designed for 1550-nm operation under $1061$-$\nano\meter$ PL illumination and monitor their characteristics at their operating wavelength.

Our study highlights a potential loophole in QKD systems: pulsed LDAs at non-operating wavelengths may compromise QKD security at mean power levels orders of magnitude below cw thresholds. Specifically, by using a $1061$-$\nano\meter$ PL with picosecond pulses, we achieve isolation reduction at~$17~\milli\watt$ comparable to that under LDA by $1$-$\watt$ PL in multi-pulse mode~\cite{ruzhitskaya2021} or several-watt 1550~$\nano\meter$ cw laser~\cite{ponosova2022}. In addition, we report a permanent isolation reduction that persists after illumination by sub-nanosecond pulses, while the isolators maintain their forward transparency. Such behavior of components has not been observed in previous studies of LDA by $1550$-$\nano\meter$ cw lasers and it may introduce new opportunities for the eavesdropper, warranting further investigation.

The pulsed fiber laser system we use is developed in the Laboratory of Active Media for Solid-State Lasers of the General Physics Institute of the Russian Academy of Sciences~\cite{zhluktova2020}. To achieve a wide range of pulse emission output characteristics, it is configured by assembling lab-made fiber devices including a master oscillator (MO), ytterbium-doped fiber power amplifiers (YDFAs), and a chirped fiber Bragg grating (CFBG) for pulse compression. \Cref{fig:PL} demonstrates PL configurations and their output characteristics---spectra and pulse shapes. The output spectra are measured using an optical spectrum analyzer (Hewlett-Packard 70950A) with a resolution of $0.1~\nano\meter$, and pulse envelope durations are obtained using a photodetector (Thorlabs DX25CF, $25~\giga\hertz$ bandwidth) with an oscilloscope (Tektronix DPO75004SX, $30~\giga\hertz$ bandwidth). The average output power is measured with an optical power meter (Ophir Orion/TH with a thermal sensor Ophir 3A).

An all-fiber ytterbium-doped (Yb) ring-cavity laser that emits at a wavelength of $1061~\nano\meter$ is used as MO~\cite{zhluktova2020}. It operates in a passive mode-locking regime based on nonlinear polarization evolution. MO provides low-power pulses with a fundamental pulse repetition rate of $1~\mega\hertz$, which corresponds to its cavity length. 

High output power is obtained owing to amplification of master pulses by a single [\cref{fig:PL}(a)] or two YDFA stages [\cref{fig:PL}(b)]. The former PL configuration yields average output power up to $800~\milli\watt$ and the latter up to $1300~\milli\watt$. However, in the case of two YDFA stages, an increase in output power broadens the spectrum and distorts the pulse shape. The pulse duration in both PL configurations falls within the sub-nanosecond range, varying from $200$ to $400~\pico\second$ depending on the gain. 

To obtain the picosecond pulse duration, a pulse compressor based on CFBG is used. As shown in the spectrum inset in~\cref{fig:PL}(c), it has a reflectivity $R$ of more than $99.9\%$ between $1060$ and $1067~\nano\meter$, which matches the main peak of the dissipative soliton of MO. The Bragg grating not only compensates for pulse chirp but also filters out emission beyond its reflection band, as can be seen in the output spectrum in \cref{fig:PL}(c). The duration of the pulse envelope is $30~\pico\second$, which corresponds to a photodetector response limit [rightmost subplot in~\cref{fig:PL}(c)]. Therefore, we analyze the duration of the compressed pulse using an autocorrelation trace measured by an autocorrelator (Femtochrome Research FR-103XL) with a resolution of $0.3~\pico\second$ (for more details of the method, see~\cite{armstrong1967, rosenfeld2018}). The autocorrelation trace (ACT), shown in the inset in~\cref{fig:PL}(c), demonstrates a coherent structure consisting of several pulses with a duration of less than $5~\pico\second$. The average output power of the compressed pulses is $17~\milli\watt$. It is limited by the damage threshold of the CFBG assembly.

\begin{figure}
	\includegraphics{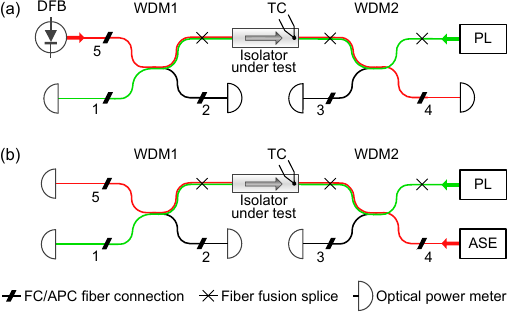}
	\caption{Experimental setup. (a)~Measurement of insertion loss at $1550~\nano\meter$. (b)~Measurement of isolation at $1550~\nano\meter$. DFB, distributed-feedback laser diode; WDM, $1061/1550~\nano\meter$ wavelength-division multiplexer; PL, pulsed laser; ASE, $1550~\nano\meter$ amplified spontaneous emission source; TC, thermocouple, fixed on the isolator housing. Light paths with minimum loss at $1061~\nano\meter$ and $1550~\nano\meter$ are shown in green (light grey) and red (dark grey).}
  \label{fig:setup}
\end{figure}

Our experimental setup is shown in \cref{fig:setup}. It has two measurement configurations. The first one enables the exposure of a fiber-optic isolator to $1061~\nano\meter$ pulsed laser radiation and the simultaneous measurement of insertion loss at the isolator's operating wavelength of $1550~\nano\meter$, while the second one provides measurement of the isolation under the same conditions. To provide efficient (de-)multiplexing of laser light at $1061~\nano\meter$ and $1550~\nano\meter$, two identical wavelength division multiplexers (WDM1 and WDM2, ordinary commercial off-the-shelf products) are placed at both sides of the tested isolator.

\begin{table*}
	\vspace{-0.7em} 
	\caption{Testing results of fiber-optic isolators. All measurements are at $1550~\nano\meter$.}
	\label{tab:ISO_result}
	\resizebox{\linewidth}{!}{
	\begin{tabular}[t]{@{\extracolsep{1ex}}c@{}c@{}c@{}c@{}c@{}c@{}c@{}c@{}c@{}c@{}c@{}c@{}}
		\hline\hline
		\multirow{3}{*}{Sample}
        &\multicolumn{2}{c}{\makecell{PL parameters }} 
		& \multicolumn{2}{c}{Initial} 
		& \multicolumn{3}{c}{Under exposure to PL}
        & \multicolumn{2}{c}{\makecell{Immediately after\\ exposure}}
        & \multicolumn{2}{c}{\makecell{1--3 days after\\ exposure}}
		\\
		\cline{2-3}
		\cline{4-5}
		\cline{6-8}
        \cline{9-10}
        \cline{11-12}
		&\makecell{Duration \\ ($\pico\second$)} &\makecell{Power\\ ($\milli\watt$)} 
        & \makecell{Isolation \\($\deci\bel$)} 
        &  \makecell{Insertion\\ loss ($\deci\bel$)}  &\makecell{Change of\\ isolation ($\deci\bel$)} &\makecell{Insertion\\ loss ($\deci\bel$)}
        &\makecell{T \\ ($\celsius$)}
        &\makecell{Change of\\ isolation ($\deci\bel$)} &\makecell{Insertion\\ loss ($\deci\bel$)}  
        &\makecell{Change of\\ isolation ($\deci\bel$)} &\makecell{Insertion\\ loss ($\deci\bel$)}\\
		\hline
		1	&360		&1160 &64.2	& 0.65  &$-36.2$	&20.0	&not measured	&$-35.1$		&8.1				&$-13.6$		&2.3 				\\ 
		2	&$<30$ 	&17		&65.4	& 0.60	&$-19.5$	&0.4	&24.9					&unchanged	&unchanged	&unchanged	&unchanged	\\ 
		2	&360		&1160	&65.4	& 0.60	&$-34.2$	&18.6	&55.3					&$-30.4$		&7.4				&$-31.7$		&15.4 			\\ 
		\hline\hline
	\end{tabular}}
	\label{tab:all}
\end{table*}

The emission of the $1061$-nm PL is injected into the tested isolator in its backward direction through WDM2. The insertion loss of WDM2 differs for PL configurations and output power owing to different PL output spectra (central plots in \cref{fig:PL}). Therefore, prior to the tests, we characterized the splitting ratio of WDM2 for each PL configuration. Then, the illumination power is set according to the calibration curves and controlled by monitoring the power at port~3 of the experimental setup. The maximum average power of sub-nanosecond pulses reaching the sample is $1160~\milli\watt$, while in the case of a picosecond pulse duration, it is about $17~\milli\watt$.

For insertion loss measurement [\cref{fig:setup}(a)], a 1550-nm cw distributed-feedback laser diode (DFB, Gooch \& Housego AA1406) with power of $46~\milli\watt$ is connected to port~5. The power transmitted through the tested isolator is monitored at port~4 of the experimental setup with an optical power meter (Thorlabs S154C). For isolation measurement [\cref{fig:setup}(b)], we connect a lab-made 1550-nm cw fiber amplified spontaneous emission source (ASE) with a power of $500~\milli\watt$ to port~4 and measure transmitted light at port~5. The isolation and insertion loss are determined by comparing input and output powers, taking into account loss in the WDMs. 

In addition, we check the sample's temperature using a thermometer (Center 301) with a Cr/Al thermocouple placed on the isolator surface.

To begin the experiment, we first measure the initial isolation and insertion loss of the sample before illuminating it with PL. We reconnect the equipment for each subsequent measurement, ensuring that the setup's components at fiber ports~4 and 5 are connected by FC/APC connectors via FC adapters that enable quick reconfiguration of the setup. 

Next, we expose the sample to a constant average power level using the configuration in~\cref{fig:setup}(a). During this illumination, we monitor the insertion loss at $1550~\nano\meter$ and the sample surface temperature until both stabilize, which takes $400~\second$ at least. Then we reconnect the equipment at ports~4 and 5 into configuration in~\cref{fig:setup}(b) to measure isolation, keeping the PL active to prevent the recovery of the sample's characteristics. Isolation is continuously measured both during and after exposure to PL. 
We note that we may not obtain isolation at $1550~\nano\meter$ during PL exposure if the isolation reduction at this wavelength is too small. In such instances, the power noise from PL pulses at port~5 exceeds the power of transmitted light at $1550~\nano\meter$. Therefore, isolation at $1550~\nano\meter$ is measured promptly after switching off PL.

After each round of illumination, we measure the insertion loss of the sample under test again. After it recovers to the initial value within the measurement accuracy of the setup, we repeat the test with a higher PL power. We stop experiments when the sample incurs irreversible damage.

This study considers the vulnerability of fiber-optic isolators in QKD systems to pulsed laser damage attacks at $1061~\nano\meter$. Fiber-optic circulators, which are also commonly used at the output of QKD sources~\cite{lucio-martinez2009,tang2014,wang2016,xia2019,liu2019}, were not tested in this study. According to previous studies~\cite{ponosova2022, tan2022}, they exhibit behavior similar to isolators owing to comparable internal design and operating principles.

\begin{figure}
	\includegraphics{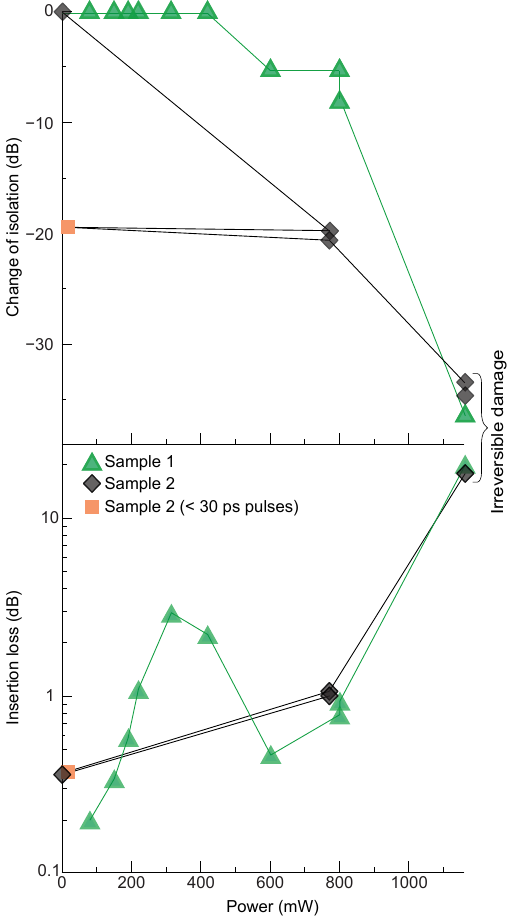}
  \caption{Isolation decrease and insertion loss of the tested samples at $1550~\nano\meter$ under illumination by the pulsed laser. At power below $600~\milli\watt$, the increase in insertion loss of sample~1 exhibits variations due to imperfect repeatability of FC/APC connection. 
Lines connect the data points in the order they were collected.}
  \label{fig:iso_under_ex}
\end{figure}

The summary of experimental results is presented in \cref{tab:all} and \cref{fig:iso_under_ex}. We test two samples of the same polarization-insensitive fiber-optic isolator model from a commercial QKD system. According to the specification, this model provides minimum guaranteed isolation of $55~\deci\bel$ at $1550~\nano\meter$ within its operating power (not exceeding $500~\milli\watt$) and temperature (from $-5$ to $+70~\celsius$).

The test procedure for the two samples differs in the sequence of applied powers. For sample~1, the average PL power is gradually increased until permanent damage occurs. Sample~2 is sequentially exposed to three distinct power regimes: $800~\milli\watt$ sub-nanosecond pulses emitted by the single-YDFA scheme~[\cref{fig:PL}(a)], followed by $17~\milli\watt$ picosecond laser pulses emitted by the pulse-compression scheme~[\cref{fig:PL}(c)], and finally $1160~\milli\watt$ sub-nanosecond pulses emitted by the double-YDFA scheme~[\cref{fig:PL}(b)].

\Cref{fig:iso_under_ex} demonstrates isolation decrease and corresponding insertion loss at $1550~\nano\meter$ under PL illumination, depending on the injected average power. Isolation decrease is calculated with reference to the initial experimental value, which is above the guaranteed level~(\cref{tab:ISO_result}).

Following \cref{fig:iso_under_ex}, under sub-nanosecond PL illumination, the degradation of isolation starts at around $600~\milli\watt$. At power levels of $770$--$800~\milli\watt$, sample~1 shows a $5$--$8~\deci\bel$ reduction in isolation, while sample~2 exhibits a more substantial $20~\deci\bel$ decrease (\cref{fig:iso_under_ex}). At $1160~\milli\watt$, both samples experience irreversible degradation, with isolation reducing by $34.2$--$36.2~\deci\bel$ and insertion loss increasing to $18.6$--$20~\deci\bel$ (\cref{tab:all}). Although both samples exhibit a comparable reduction in isolation during illumination, they recover differently after exposure. Isolation of sample~1 partially recovers to $50.6~\deci\bel$  within 1--3 days, while isolation of sample~2 remains at $33.7~\deci\bel$~(\cref{tab:ISO_result}). This difference may arise from exposing the latter to the maximum PL power twice. Consequently, both samples exhibit an irreversible laser-induced isolation reduction while preserving forward transmissivity. Note that in all tests, the surface temperature does not exceed $+58~\celsius$ (\cref{tab:all}), remaining within the specified operating range from $-5$ to $+70~\celsius$. The negligible temperature increase and low isolation of about $15~\deci\bel$ at $1061~\nano\meter$ observed in another experiment~\cite{ruzhitskaya2021} suggest that non-thermal mechanisms are responsible for the damage, in contrast to previously reported cw LDA effects at $1550~\nano\meter$~\cite{ponosova2022}.

We conduct a test on sample~2 under illumination by picosecond pulses following its exposure to $770~\milli\watt$ sub-nanosecond pulses. Prior to this test, the sample's characteristics had returned to their initial values. However, we remain open to the possibility that pre-exposure may have influenced the test results.
The sample~2 exhibits an isolation reduction of $19.5~\deci\bel$ when illuminated with $17~\milli\watt$ picosecond pulses. In contrast, achieving the same isolation decrease with sub-nanosecond pulses required an average power of $770~\milli\watt$ (see \cref{fig:iso_under_ex}). This $45$-fold decrease in average laser power required to change the characteristics of the isolator underscores the effectiveness of short pulses for LDAs owing to their high peak intensity. Such low-average-power attacks might avoid detection by countermeasures designed for cw or high-power threats. This finding necessitates additional experimental validation, also at a higher pulse energy, as the observed effects may depend on manufacturing variations. We could not repeat the test timely on more samples as this was our last available sample.

\begin{figure}
	\includegraphics{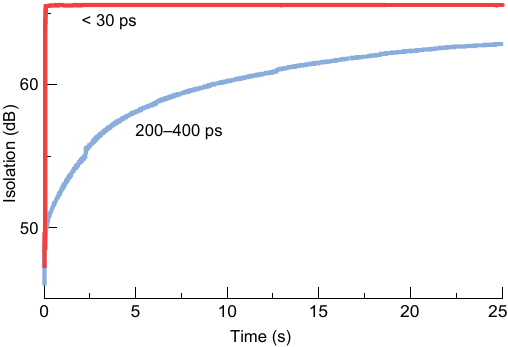}
	\caption{Isolation recovery of sample~2 immediately after illumination by $770~\milli\watt$ sub-nanosecond pulses and $17~\milli\watt$ picosecond pulses. PL is switched off at $0~\second$.  
}
	\label{fig:recovery}
\end{figure}

\Cref{fig:recovery} demonstrates the short-term isolation recovery dynamics of sample~2 over the first $25~\second$ after exposure to $17~\milli\watt$ picosecond pulses and $770~\milli\watt$ sub-nanosecond pulses. Despite comparable isolation reductions for these exposure regimes, sample recovery times differ. After exposure to picosecond pulses, isolation recovers immediately, faster than the $10$-$\milli\second$ temporal resolution of OPM. In contrast, after sub-nanosecond pulse exposure, isolation recovers more gradually, reaching  $62.8~\deci\bel$ within the first $25~\second$. We continued monitoring the long-term recovery of the latter outcome and observed that the isolation recovered to $64.5~\deci\bel$ within an hour and to the initial level after one day. This difference hints at different damage mechanisms of picosecond and sub-nanosecond pulses at the same wavelength of $1061~\nano\meter$.

We see that, unlike cw LDA that often destroys the quantum channel~\cite{ponosova2022}, LDA using PL reduces isolation while preserving transmission in the forward direction. 
An eavesdropper could potentially exploit this effect to carry out attacks. However, the strategy and efficiency of the attack significantly depend on the practical implementation of the QKD system. Here, we discuss two common cases of transmitter design and threats to their security in the context of pulsed LDA. If the affected isolator is counted in secure key rate estimation in the presence of the Trojan-horse attack, the reduction of its isolation, either temporary or permanent, will significantly reduce the maximum secure key generation distance~\cite{lucamarini2015,tan2025}. In another scenario, Alice might implement a countermeasure consisting of an additional fiber-optic isolator not accounted for in the security proofs of the QKD system, as suggested in our previous study~\cite{ponosova2022}. In this case, the effectiveness of pulsed LDA relies on the damage thresholds of the components located behind the last isolator, along with the maximum transmitted power allowed through it. Unfortunately, our current limitations in the maximum available power of lasers prevent us from experimentally verifying this latter scenario. This remains a topic for future research. Our results however show that existing QKD component testing standards are insufficient to guarantee protection against different scenarios of LDA.

In conclusion, we have studied the effect of $1061$-$\nano\meter$ pulsed lasers on fiber-optic isolators. While our test is not a full eavesdropping attack, it is the necessary first step in analysing the potential vulnerability. The isolation reduction by about $20~\deci\bel$ under exposure to PL is demonstrated using picosecond laser pulses with average power of just $17~\milli\watt$. Although this result was obtained on a single sample, we consider the single experimental point to be a significant preliminary finding. It warrants a further investigation. We also induce a permanent isolation decrease without breaking the quantum channel by illuminating the isolators with sub-nanosecond pulses. Both effects are not properly accounted for in the current QKD security analysis.

Generally, our results highlight that the possibilities of LDA may be extended by using a wide range of laser radiation parameters. Failure to address this residual vulnerability may leave practical QKD implementations vulnerable. Our experimental study on a specific countermeasure, fiber-optic isolators, indicates that such measures may not effectively mitigate the light-injection attacks by pulsed lasers. While test the component at a single attack wavelength, the eavesdropper is free to select any one. Thus, to reliably assert the effectiveness of various countermeasures against pulsed LDA scenarios in a wide spectral range, further experimental testing under different laser operation regimes and wavelengths, and incorporating its results into the security model, are essential.

\bigskip

This work was funded by the Russian Science Foundation (grant 23-79-30017). D.T.\ and V.M.\ acknowledge funding from the Galician Regional Government (consolidation of research units: atlanTTic and own funding through the ``Planes Complementarios de I+D+I con las Comunidades Autonomas'' in Quantum Communication), MICIN with funding from the European Union NextGenerationEU (PRTR-C17.I1), the ``Hub Nacional de Excelencia en Comunicaciones Cu{\' a}nticas'' funded by the Spanish Ministry for Digital Transformation and the Public Service and the European Union NextGenerationEU, and the European Union's Horizon Europe Framework Programme under Marie Sk\l{}odowska-Curie grant 101072637 (project QSI) and project ``Quantum Security Networks Partnership'' (QSNP; grant 101114043). A.H.\ acknowledges funding from the National Natural Science Foundation of China (grant 62371459).

\section*{Author declarations}

\textbf{Conflict of interest.} The authors have no conflicts to disclose.

\medskip
\textbf{Author contributions.}
	I.Z.,\ D.R.,\ A.P.,\ and V.K.\ conducted the experiment. A.W.\ manufactured CFBG for pulse compression. A.H.\ provided samples for testing. A.P.,\ D.R.,\ I.Z.,\ and D.T.\ analysed the data and wrote the article with help from all authors. V.K.,\ V.T.,\ and V.M.\ supervised the project.  

\medskip
\textbf{Data availability.} The data that support the findings of this study are available from the corresponding author upon reasonable request.

\def\bibsection{\bigskip\bigskip} 
\bibliography{library}

\end{document}